\documentclass[12pt,letterpaper]{article}
\usepackage[a4paper, total={7in, 10in}]{geometry}

\usepackage{graphicx}
\usepackage{helvet}
\usepackage{authblk}
\usepackage{hyperref}
\usepackage{amsmath} 
\usepackage{amssymb} 
\usepackage{orcidlink} 
\usepackage[super,comma,sort&compress]  
   {natbib}
\usepackage[right]{lineno} \linenumbers

\makeatletter
\renewcommand{\maketitle}{\bgroup\setlength{\parindent}{0pt}
\begin{flushleft}
  \textbf{\@title}
  
  \@author
\end{flushleft}\egroup}
\makeatother

\title{Toward Carnot efficient high output power heat engines using bubbly two-phase flow}
\date{}

\author[1,3,4,*,\orcidlink{0000-0002-9006-6665}]{Dror Miron}
\author[1,3,4]{Yuval Neumann}
\author[3,4]{Joseph Cassell}
\author[2,3]{Nir Feintuch}
\author[3]{Alexey Shinkarenko}
\author[1,2,3,**]{Carmel Rotschild}

\affil[1]{The Nancy and Stephen Grand Technion Energy Program, Technion – Israel Institute of Technology, Haifa 3200003, Israel}
\affil[2]{Faculty of Mechanical Engineering, Technion – Israel Institute of Technology, Haifa 3200003, Israel}
\affil[3]{Lava Energy LTD.}

\affil[*]{Correspondence: sdmiron@campus.technion.ac.il}
\affil[**]{Correspondence: carmelr@technion.ac.il}
\affil[4]{These authors contributed equally.}

\begin{document}

\maketitle

\section*{SUMMARY}

Thermodynamic gas power cycles achieving Carnot efficiency require isothermal expansion, which is associated with slow processes and results in negligible power output. This study proposes a practical method for rapid near-isothermal gas expansion, facilitating efficient heat engines without sacrificing power. The method involves bubble expansion within a heat transfer liquid, ensuring efficient and near-isothermal heat exchange. The mixture is accelerated through a converging-diverging nozzle, converting thermal energy into kinetic energy, thereby rotating a reaction turbine for electricity generation.
Experiments with air and water yielded a polytropic index $<$1.052, enabling up to 71\% more work extraction than adiabatic expansion. Simulations indicate that utilizing these nozzles for thrust generation decreases heat transfer irreversibilities in the heat engine, resulting in up to 22.6\% higher power output than an ideal heat engine based on the organic Rankine cycle. This work paves the way for an efficient and high-power heat-to-power solution.


\section*{Graphical abstract}
\noindent\includegraphics[width=1\linewidth]{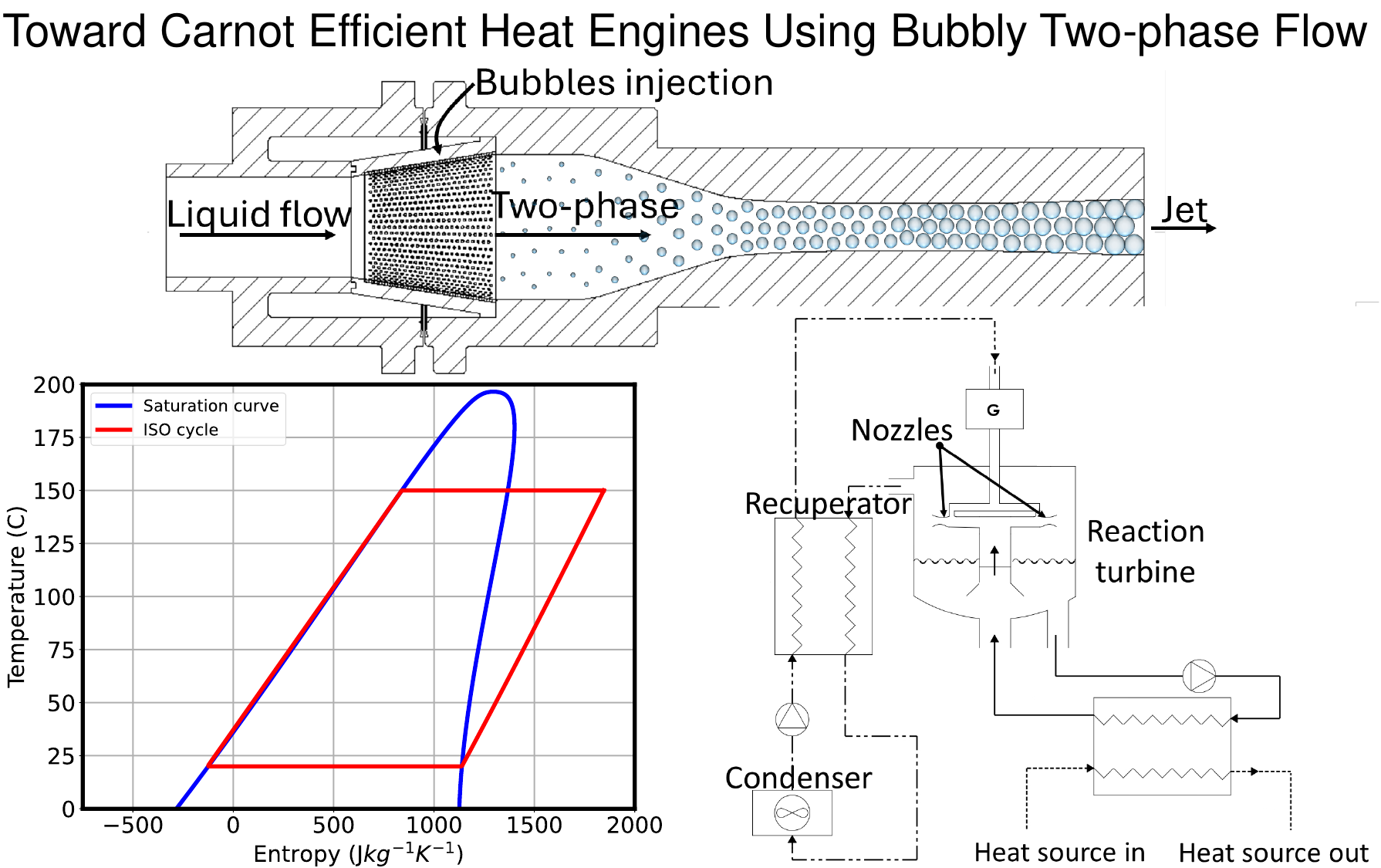}

\section*{KEYWORDS}


Isothermal expansion, Heat engine, Waste heat recovery, Carnot efficiency, Two-phase nozzle

\section*{INTRODUCTION}

To limit global warming corresponding to international goals, global greenhouse gas emissions should be urgently reduced \cite{Masson-Delmotte2018}. One of the promising methods to limit greenhouse gas emissions is by improving energy efficiency. As 72\% of global primary energy production is lost to waste heat \cite{Forman2016}, there is great potential for energy to be sourced from waste heat recovery methods. 
Much of this potential lies in industrial applications, where heat is dissipated as a by-product in tens to hundreds of kilowatts, for which available heat engines are inefficient and expensive. 

Due to its enhanced thermodynamic performance, power systems based on the organic Rankine cycle (ORC) have been considered promising for waste heat \cite{geffroy2021, Bianchi2011} and geothermal energy recovery systems \cite{GUZOVIC2010}. Therefore, improving ORC systems' performance received research attention in recent years \cite{ZHAI2016, ZIVIANI2018, LI2021, Castelli2019, NEMATI2017, WANG2013}.
Specifically, it has been suggested to implement quasi-isothermal expansion as an alternative to adiabatic expansion since more work can be extracted \cite{Igobo2014, ZIVIANI2018}. For example, studies indicated a flooded-expansion method to achieve quasi-isothermal expansion \cite{woodland2014, WOODLAND2013, ZIVIANI2018}. Such cycles involve flooding the expansion device (screw or scroll expander) with a liquid that is in thermal equilibrium with the working fluid (WF). The liquid acts as a heat reservoir for the expanding vapor, which allows it to maintain its temperature during the expansion process. The isothermal expansion also produces super-heated vapor at the expander outlet, facilitating internal regeneration normally unavailable in conventional ORC \cite{WOODLAND2013}. Although promising, studies investigating flooding the screw expander found no change to the outlet temperature (i.e., isothermal expansion has not been achieved) and a decreased trend of the cycle efficiency when increasing the flooding ratio\cite{Nieuwenhuyse2020, ZIVIANI2018, LI2021}. 
The scroll expander showed deteriorating performance when increasing the flooding ratio due to pressure drop in the scroll expander inlet \cite{Igobo2014,lemort2008}. Another challenge mentioned in this configuration is phase separation after the expander. Consequently, a practical approach has yet to be known for realizing quasi-isothermal expansion in organic cycles.

In contrast to expander flooding, studies in marine propulsion have proposed employing air bubble expansion within water to achieve isothermal expansion. This method significantly increases the contact area between the fluids and facilitates efficient heat transfer. Additionally, this method substantially elevates the liquid-to-gas mass ratio relative to the liquid flooding technique, promoting gas expansion within a heat reservoir. Integrating these effects in a converging nozzle has enhanced thrust by allowing more effective work extraction compared to single-phase air ramjets \cite{FU2009, Zhang2018, Gany2018}. To further augment thrust, it has been proposed that choked flow conditions be attained. As such, a converging-diverging nozzle has been studied and experimented with, demonstrating supersonic flow and thrust enhancement \cite{wu2015}.


This study presents the development of an efficient, high-power heat engine based on a novel thermodynamic vapor cycle incorporating quasi-isothermal expansion of bubbles in a heat transfer liquid (HTL). The large bubbles' surface area supports efficient heat transfer between the phases, and
the substantial mass and heat capacity discrepancies between the phases facilitate minimal temperature drop, thereby implementing a quasi-isothermal process. The two-phase mixture flows through converging-diverging nozzles, converting the thermal energy to thrust, which rotates a reaction turbine. 
To this end, we combine experimental pressure measurements and computational fluid dynamic (CFD) analysis of a two-phase flow in a converging-diverging nozzle to extract an upper boundary for the polytropic index of air and quantify the work enhancement of the near-isothermal compared to adiabatic expansion.
Building on this, we simulate the performance of a heat engine that integrates the nozzles as expanders to drive a reaction turbine for electricity generation from a constant-temperature heat source. 
The results highlight the potential of this design, as a polytropic index of air $n<$1.052 is determined, resulting in work extraction $>$91\% of perfect isothermal work. Thermodynamic simulations indicate that the proposed heat engine produces up to 22.6\% greater power output than an ORC-based counterpart for a 100–374 $^\circ C$ constant temperature heat source. 
Importantly, the isothermal expansion of the cycle facilitates the use of 'wet' fluids such as water—fluids that challenge the performance in ORC systems but are well-suited for high-temperature heat sources.
Since the turbine's practical efficiency is obscure, we demonstrate that turbine efficiencies between 70\% and 90\% yield an improvement of 42.7\% and 90.7\%, respectively, compared to the ORC heat engines with turbine isentropic efficiency of 75\% at the temperature of 373 $^\circ C$. These findings provide a basis for developing efficient heat engines tailored to heat-to-power applications. 



\section*{RESULTS AND DISCUSSION}
\subsection*{Nozzle design}

The nozzle is designed according to a previously developed model for two-phase flow in a converging-diverging nozzle \cite{Singh2014}. The design is based on the homogenous model of a bubbly mixture flowing in the nozzle, which was shown valid by others \cite{Albagli2003, THANG1979, Mor2004, FU2009, Zhang2018}. Under this model, we assume that the phases are in pressure, temperature, and velocity equilibrium. In contrast to the nozzles designed for underwater propulsion where the pressure at the injection chamber is dictated and limited by the cruise velocity, the inlet pressure is a free parameter controlled by a pump, which results in supersonic flow. 

To illustrate the physical principle and evaluate the expansion, we introduce air into a water stream at ambient temperature. The injection pressure at the design point was chosen to be 6 bar, an air mass flow rate of 13.1 $g/s$, an injection temperature of 26 $^\circ C$, and a water volumetric flow rate of 14.6 $m^3/hr$. The geometry of the cross-section is selected as circular. The nozzle is designed to implement a constant $dp/dt$ to ease the heat transfer between the phases. Figure 1 shows a schematic representation of the designed nozzle.

\noindent\includegraphics[width=0.85\linewidth]{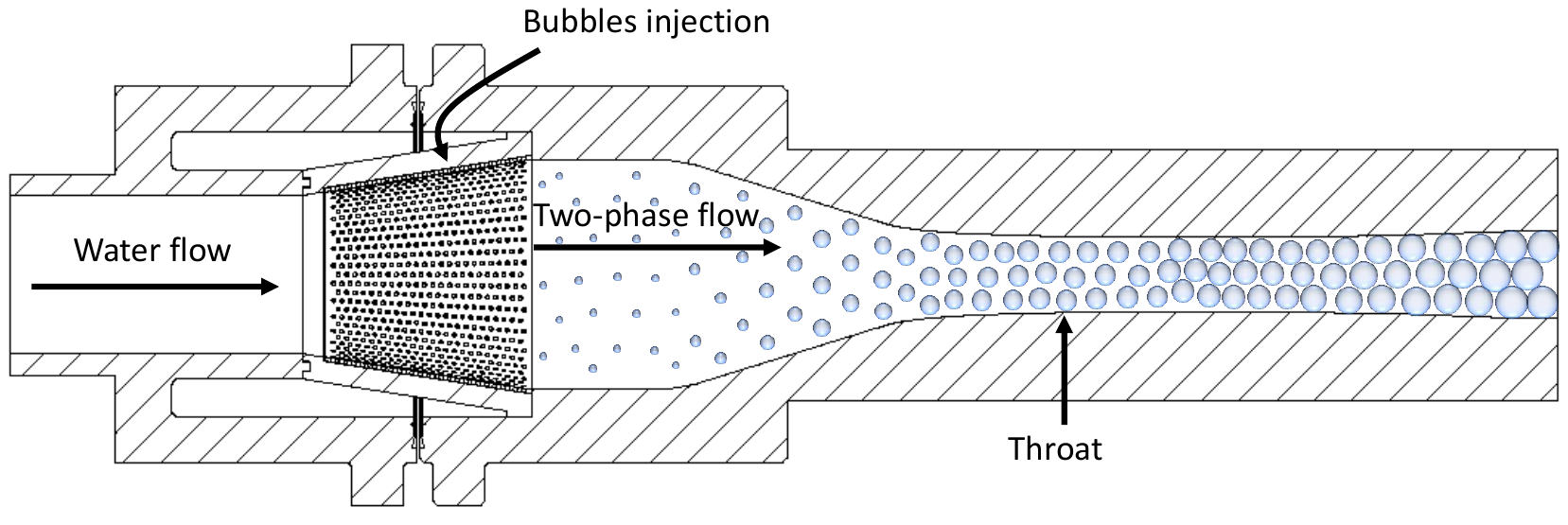}
\subsection*{Figure 1. Nozzle design drawing}

\subsection*{Near-isothermal expansion} \label{Isothermal efficiency calculation}

During the experiments, static pressure was measured at the nozzle's injection, outlet, and other locations along the nozzle.
Figure 2 presents the pressure distribution and the nozzle's diameter in the nozzle (solid black curve). It indicates that supersonic conditions are met in the nozzle since the pressure continues to drop after the throat until it reaches atmospheric pressure at the nozzle's outlet. This pressure distribution matches the nozzle's design, i.e., reducing the pressure from a 6 bar in the air injection chamber to atmospheric pressure in the nozzle's outlet. Since isothermal expansion was assumed in the nozzle's design, this is the first reinforcement of our isothermal process assumption. We note that the pressure measurement at the throat stands out, where the measured pressure is considerably lower than anticipated. We ascribe this discrepancy to the high throat sensitivity to manufacturing flaws, slip \cite{wang2002}, and tap diameter \cite{furuichi2015} due to high pressure and velocity gradients in this region. 
A computational fluid dynamics (CFD) simulation was performed to analyze the anticipated pressure distribution at the working conditions, which is detailed in the experimental procedure subsection. The CFD outcomes closely align with the experimental data, corroborating the isothermal expansion hypothesis, as the CFD model assumes uniform mixture temperature.

\noindent\includegraphics[width=0.85\linewidth]{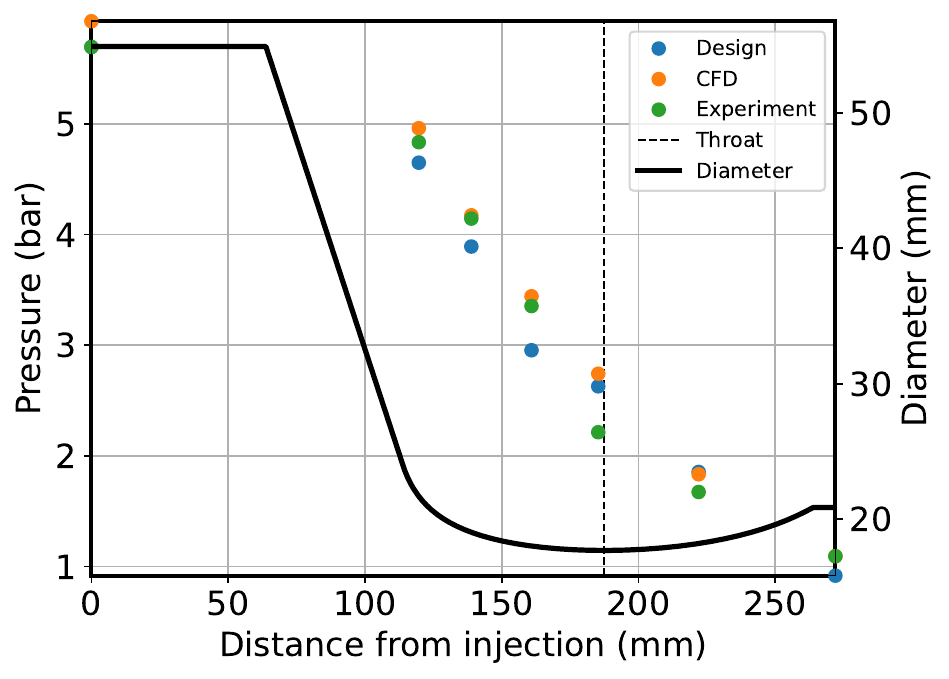}
\subsection*{Figure 2. Pressure distribution comparison between the measured, simulated, and designed values}
Subsequently, we employ the polytropic index to gauge how much this process approximates a strictly isothermal expansion.
To quantify the polytropic index, we apply the energy conservation equation to the nozzle's outlet, where the pressure is P, and the mixture's velocity is expressed by:


\begin{equation} \label{energyConservation}
u_{out}^2 = \frac{2\dot{V}_w(P_{inj}-P_{out})}{\dot{m}_w + \dot{m}_a} + u_{inj}^2 + \frac{2}{\dot{m}_w + \dot{m}_a}   \times (\dot{W}_a - \dot{E}_{loss})
\end{equation}

where $\dot{V}_w, P_{inj}, P_{out}, \dot{m}_a,\dot{m}_w,u_{inj},u_{out},\dot{W}_a$ and $\dot{E}_{loss}$ are the water volumetric flow rate, air injection pressure, outlet pressure, air mass flow rate, water mass flow rate, mixture's velocity at injection, mixture's velocity at nozzle's outlet, work rate exerted by air, and energy losses, respectively.
Further discussion on the governing equations can be found in the experimental procedure section.
The air expansion work is assumed to be a polytropic process, for which the work rate is calculated by \cite{Igobo2014}:

\begin{equation} \label{W}
\dot{W}_a = P_{inj}\dot{V}_{a,inj}\frac{1-(\frac{P_{inj}}{P_{out}})^{\frac{1-n}{n}}}{n-1}
\end{equation}
where n is the polytropic index.
As a result, equation \ref{energyConservation} involves two unknown variables: the polytropic index and the head losses in the nozzle.

To gain insights on the energy losses in the nozzle, we evaluate the expression $\dot{E}_{loss} =\dot{V}_w \Delta P$, where $\Delta P$ is the head loss, and is calculated by $\Delta P = \zeta \frac{\rho u^2}{2} $. Here, $\zeta$, $\rho$, and $u$ are the resistance coefficient, mixture's density and inlet velocity, respectively \cite{Idelchik2007}.


In the above equation, $\zeta$ depends on geometrical factors while the term $\rho u^2$ strongly depends on the void fraction and, thus, the temperature and the polytropic index. Consequently, comprehending the influence of the polytropic index on this parameter is sufficient to infer the nuzzle's head loss characteristics. This relationship is quantified by the equation below:

\begin{equation} \label{Loss expression}
\rho u^2 = \left[\alpha\rho_{a} + (1-\alpha)\rho_{w}\right] \times \frac{\dot{V}^2_{w}}{A^2(1-\alpha)^2} = \frac{\dot{V}^2_{w}}{A^2} \times \left[\frac{\alpha\rho_{a}}{(1-\alpha)^2} + \frac{\rho_{w}}{1-\alpha}\right],
\end{equation}
where $\alpha$ is the void fraction and $A$ is the cross-sectional area.
Figure 3 plots the above term as a function of the polytropic index using the measured conditions at the nozzle's outlet. 
The head losses monotonically reduce with increasing the polytropic index, indicating that the largest head losses are received for the pure isothermal case. Therefore, the head losses obtained from the CFD simulation, representing perfect heat transfer between the phases, can be considered an upper bound for the actual head loss.

\noindent\includegraphics[width=0.85\linewidth]{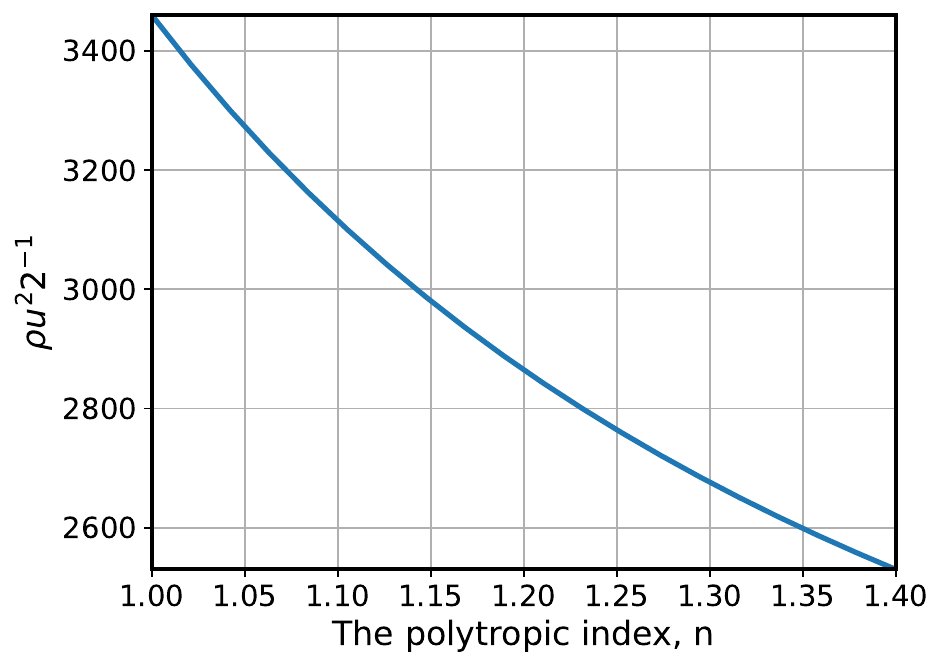}
\subsection*{Figure 3. Head loss dependence on the polytropic index}


In Figure 4, we plot the solution for the energy conservation equation (equation \ref{energyConservation}) for the nominal conditions and water volumetric flow rates error of 0.5\%. Since both head losses and $n$ are unknown, each point on this curve is a solution to this equation. Using equation 1, the head losses derived from the CFD simulation is 0.84 bar. This value is used as an upper bound to our experiment; thereby, we can extract an upper bound on the polytropic index, $n$. From this analysis, the upper bound for the polytropic index is 1.052.

\noindent\includegraphics[width=0.85\linewidth]{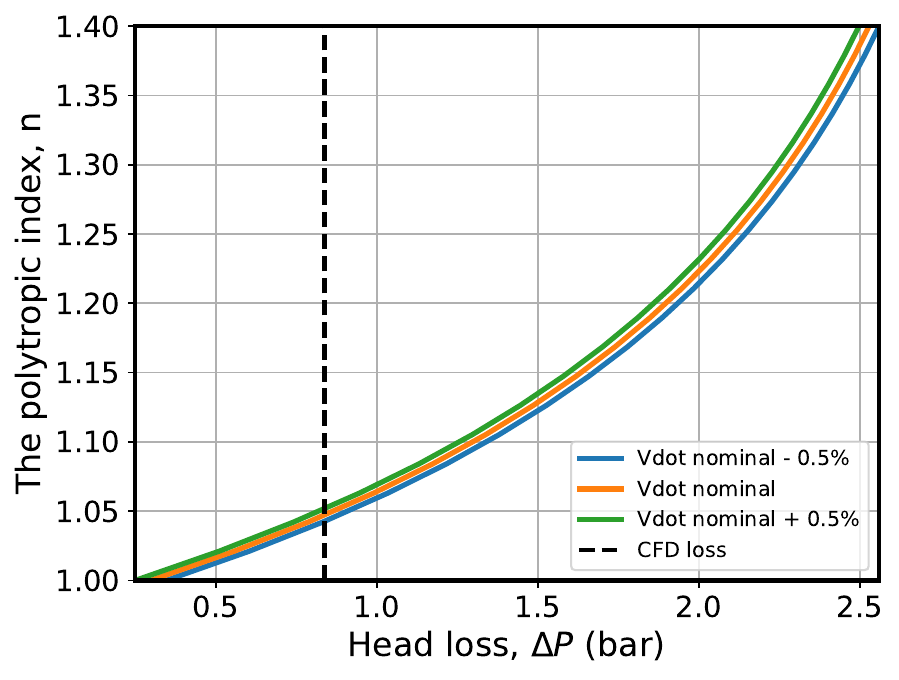}
\subsection*{Figure 4. Solution to equation \ref{energyConservation} using measured data.}

One of the benefits of isothermal expansion stems from the additional work that can be extracted in the expansion process.
Hence, it is instructing to understand the improved extracted work compared to the adiabatic process, which represents gas expansion with no heat exchange. Therefore, we define the isothermal efficiency by the ratio between the polytropic-index dependent work, $\dot{W}_a = P_{inj}\dot{V}_{a,inj}\frac{1-(\frac{P_{inj}}{P_{out}})^{\frac{1-n}{n}}}{n-1}$, and the isothermal work $\dot{W}_{iso} = P_{inj}\dot{V}_{a,inj}ln\left(\frac{P_{inj}}{P_{out}}\right)$ \cite{Igobo2014}:
\begin{equation} \label{eta_iso}
\eta_{iso} = \frac{\dot{W}_a}{\dot{W}_{iso}}=\frac{1-r_p^{\frac{1-n}{n}}}{(n-1)\mathrm{ln}(r_p)}, 
\end{equation}
where $P_{out}$ is the outlet pressure and $r_p = \frac{P_{inj}}{P_{out}}$ is the pressure ratio.

Figure 5 shows the isothermal efficiency using the extracted polytropic index of $n=1.052$ for pressure ratios up to 10, for which isothermal efficiencies $>94\%$ are attained. As such, an additional 44-71 \% more work can be extracted compared to adiabatic expansion, for which $n=1.4$. 

\noindent\includegraphics[width=0.85\linewidth]{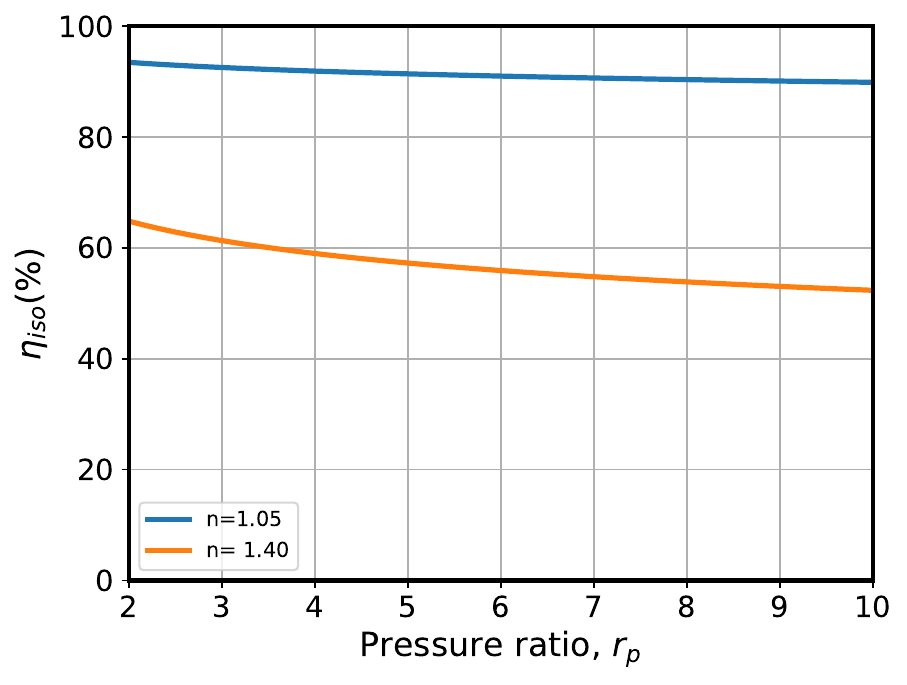}
\subsection*{Figure 5. Isothermal efficiency for varying pressure ratios.}



\subsection*{High impeller efficiency}

The near-isothermal expansion of bubbles experienced in our nozzle reveals opportunities for a new type of heat engine utilizing such nozzles for a reaction turbine. The turbine consists of an impeller connected to nozzles.
The rotor leg is submerged in the HTL reservoir, sucking HTL into the rotor. Similarly to a centrifugal pump impeller, the impeller vanes pressurize the HTL until reaching maximal static pressure at the end of the impeller, where the nozzles begin, and liquid WF is injected and mixed with the HTL. However, unlike the centrifugal pump, our impeller has no stator, minimizing energy losses while maintaining high design flexibility. In the nozzles, heat is transferred from the HTL to the WF until the WF is fully evaporated, and the mixture is accelerated until it reaches the nozzles' outlet, where the jet flows tangentially, creating a thrust force and rotates the shaft for electricity generation. An example of the impeller design is shown in Figure 6.


\noindent\includegraphics[width=0.85\linewidth]{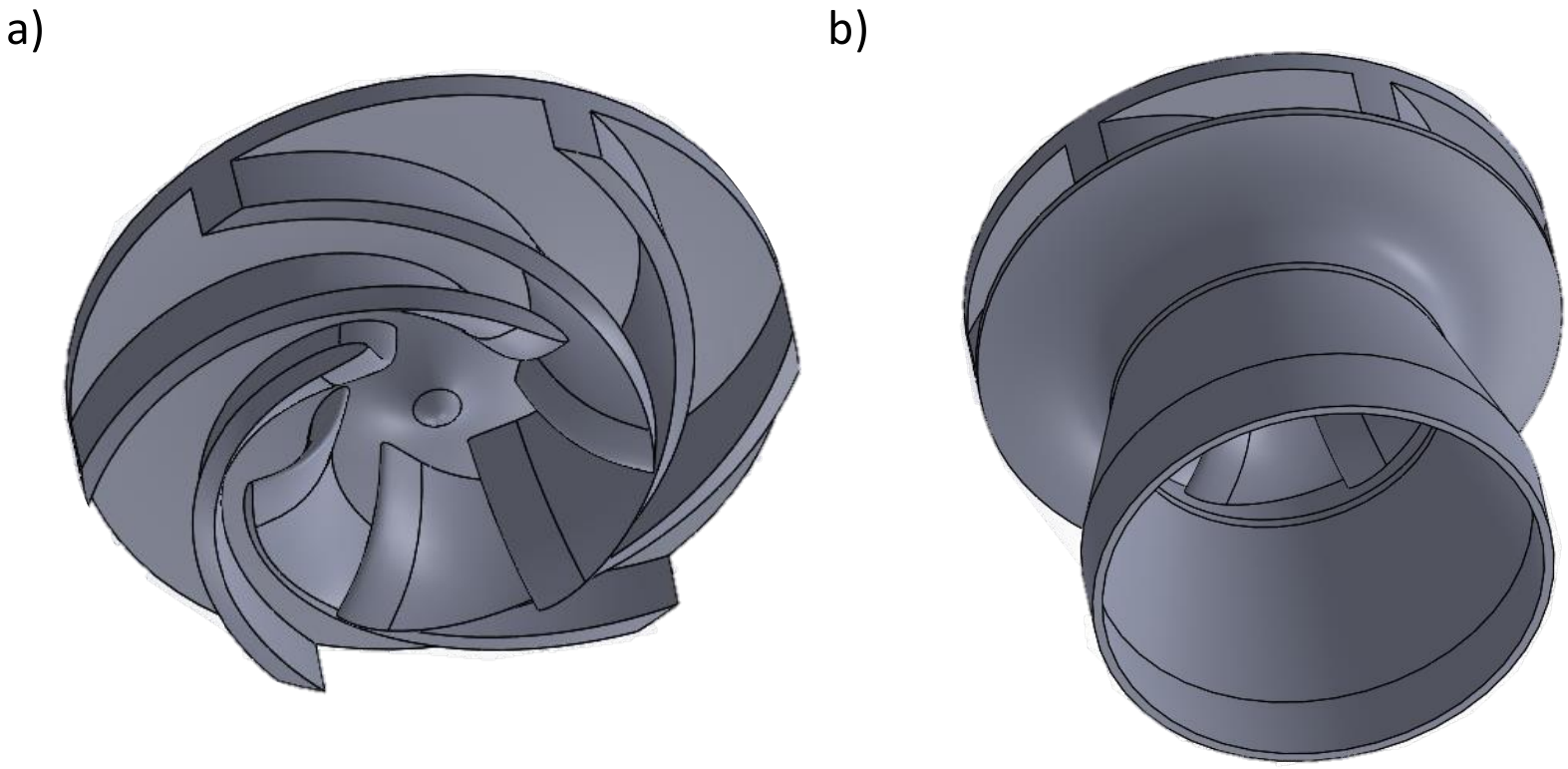}
\subsection*{Figure 6. Impeller design. (a) hidden shroud, and (b) with shroud.}



Figure 7 exemplifies CFD simulation of an impeller design for 10.5 bar injection pressure, resulting in an impeller efficiency of $\sim97\%$.

\noindent\includegraphics[width=0.85\linewidth]{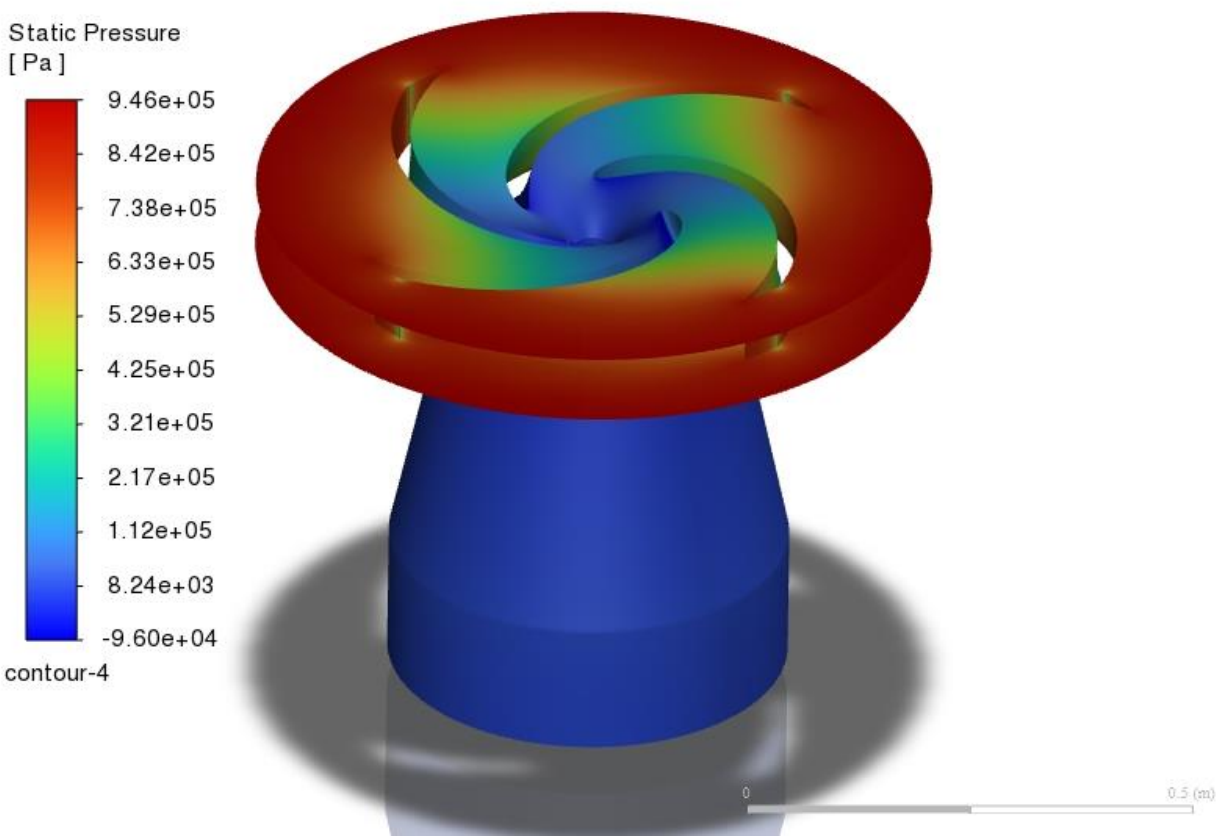}
\subsection*{Figure 7. CFD results of the impeller at 1500 RPM. Pressure values are expressed in bar gauge.}



\subsection*{Description of the ISO cycle}

This subsection details the novel thermodynamic cycle (ISO cycle henceforth). In this cycle, depicted in Figure 8, the HTL transfers thermal energy from an outside heat source to the WF. Thermal energy is converted into kinetic energy at the nozzle, creating thrust and rotating a reaction turbine to generate electricity. \newline
The heat source flows through a heat exchanger (1$\rightarrow$2), transferring heat to the HTL (3$ \rightarrow$4). The HTL is pumped by the turbine in conjunction with colder HTL from the reservoir (5). The HTL flow from the reservoir is used to obtain the desired HTL flow rate in the nozzle, resulting in a specific void fraction. Before reaching the nozzle, the impeller increases the HTL's pressure to the WF's saturation pressure. The HTL reaches the nozzle's inlet and mixes with colder WF liquid (6) in the nozzle's injection chamber. Heat transfers to the WF, which raises its temperature to its evaporation temperature. The WF then completely evaporates and expands inside the nozzle within the HTL until it reaches its outlet (7). This expansion raises the mixture's velocity and generates thrust, which turns the reaction turbine. The rotor rotates in a tank that functions as a vertical vapor-liquid separator, separating the WF from the HTL after the expansion by generating a film flow of the mixture on the tank boundaries. The hot WF gas enters a recuperator, transferring heat from the hot WF to the cold condensed WF before it enters the condenser (8$\rightarrow$9). This internal regeneration, facilitated by the isothermal expansion, is used to heat the cold liquid WF, minimizing heat transfer irreversibilities. Then, the liquid WF is pumped through the recuperator (10$ \rightarrow$11). The WF is then injected into the nozzles, mixing with the hot HTL (6). 

\noindent\includegraphics[width=\linewidth]{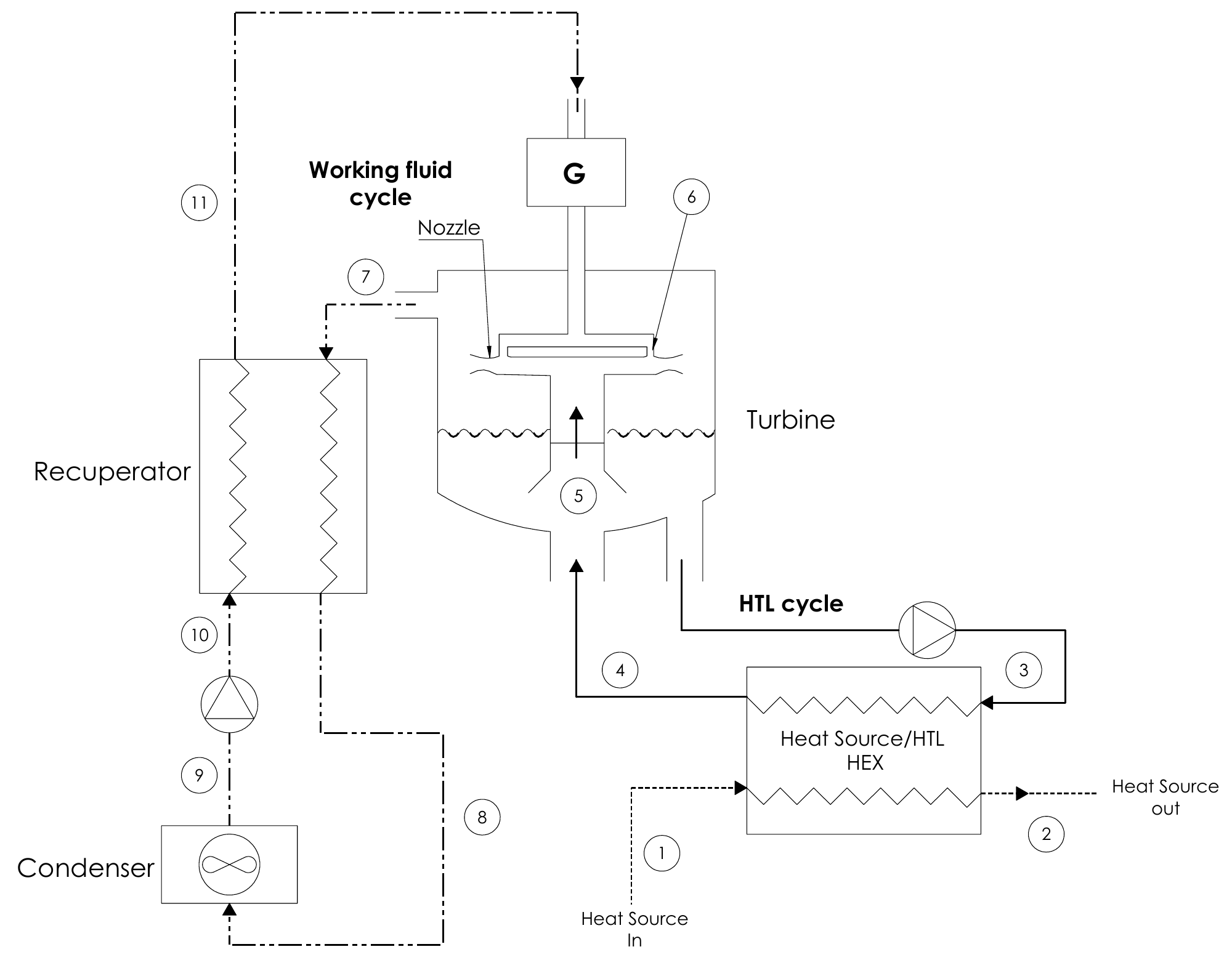}
\subsection*{Figure 8. Schematic of the ISO cycle.}
\subsection*{ISO cycle performance analysis}

In the ISO cycle, isothermal expansion drives vapors away from saturation conditions, ensuring the fluid remains in a single phase. This, in turn, enables the use of 'wet' fluids—like water—that have a higher critical temperature than the 'dry' fluids. This lies in contrast to ORC-based heat engines, where it is advisable to employ 'dry' fluids to limit superheat and sustain a dry turbine \cite{CHEN2010}.
 
Therefore, the ISO cycle was simulated with water as the WF and a heat source in 100-374 ${}^\circ$C, covering a substantial portion of the waste heat recovery temperature range. The simulated heat source is characterized by a constant temperature, representing condensation scenarios harnessing constant-temperature latent heat. This application is optimally suited for the ISO cycle, as the predominant portion of the heat involved is latent heat, allowing significant heat extraction with minimal temperature change. The scale of the system is a few hundred kW, which is the scale needed for many industrial waste heat recovery systems. The heat sink temperature was chosen as 20 ${}^\circ$C. 


To understand the proposed cycle's benefits, we compare its performance to a conventional recuperated ORC. In the simulated ORC cycle, the WF is heated in a boiler by the same heat source and expands in a turbine. Then, the colder WF transfers heat by a recuperator before it condenses and is pumped back into the boiler through the recuperator. In the two cycles, the temperature of the hot WF is optimized to extract maximum power. 

The ORC is simulated with cyclopentane as the WF using the same heat source. Cyclopentane is chosen due to its high critical and working temperatures of 238.54${}^\circ$C and 300${}^\circ$C, respectively \cite{INVERNIZZI2017}. Furthermore, cyclopentane is suggested as a dry fluid for the organic Rankine cycle for low-grade heat conversion \cite{CHEN2010}. The thermophysical properties of the materials were acquired from the National Institute of Standards and Technology (NIST) REFPROP database \cite{Nist}.

Figure 9 exemplifies T-S diagrams of the two cycles using heat source temperature ($T_H$) of 200 ${}^\circ$C with cyclopentane (a) and water (b). The larger ISO cycle in Figure 9 a implies that more output power is available than the conventional ORC per WF unit mass flow. The isothermal expansion allows a more significant internal regeneration using a recuperator (289 VS 39 kJ/kg), and as such, the liquid preheating portion is reduced, and more heat is transferred isothermally to the WF. Consequently, heat transfer irreversibilities are reduced compared to the ORC, resulting in increased thermal efficiency. Additionally, no superheating is needed in the ISO case, further reducing irreversibilities and increasing the turbine's pressure ratio.

\noindent\includegraphics[width=\linewidth]{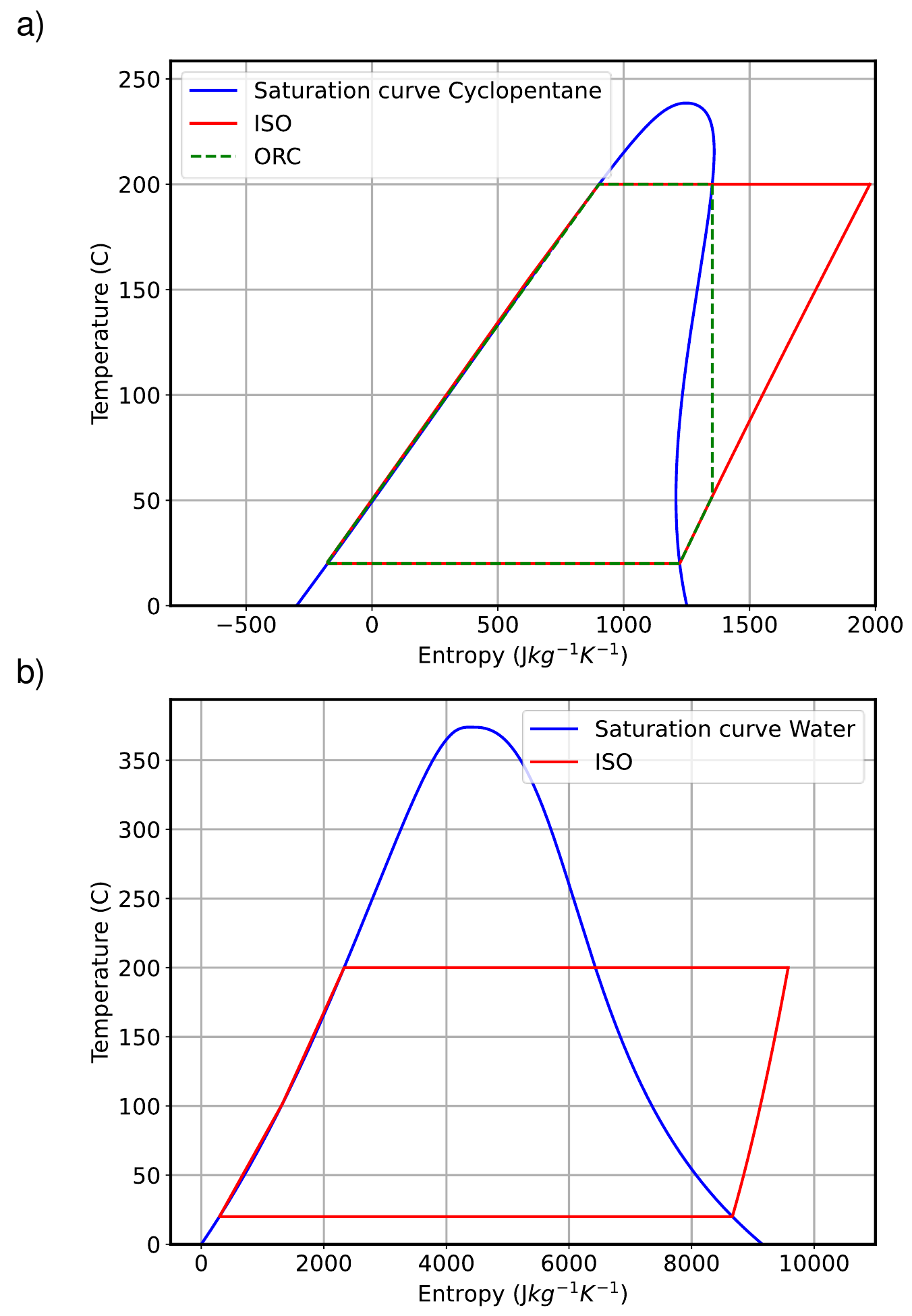}
\subsection*{Figure 9. T-S diagrams of the compared cycles at $T_H = 200 ^\circ C$. a) Cyclopentane comparison. b) ISO cycle with water as the WF.}

To understand the performance potential of the ISO cycle, we compare the efficiency and power output of an ideal ISO cycle (using both water and cyclopentane as the WF) to an ideal organic Rankine cycle using a constant-temperature heat source, considering 1 MW heat input. Figures 10 a and b show that the ISO cycle outperforms ORC in efficiency and power output metrics, respectively. The gains in power output are tabulated in Table 1, indicating that the ISO cycle can extract up to 22.6\% more power than the ideal ORC in the analyzed temperature range. 
In terms of the specific work, calculated by dividing the output power by the WF's mass flow rate, the improvement is more profound, reaching a 7-fold improvement compared to the ORC. This stems from the high specific heat possible with water compared to cyclopentane.
Figure 10 a also shows the Carnot efficiency, which is calculated as $\eta_C = 1 - \frac{T_C}{T_H}$.
The second law efficiency of the proposed cycle, calculated by the cycle's efficiency divided by the Carnot efficiency, is 93.8\%-97.3\%, approaching the Carnot efficiency, compared to $<$87\% using the ORC. The dips in the ORC power and efficiency curves at the temperature of $T_H=$240$^\circ C$ are attributed to the transition between operating without superheating at the lower temperatures to the higher temperatures for which superheating is required. This phenomenon is also visible, albeit less significant, in the ISO cycle when using cyclopentane.

\noindent\includegraphics[width=\linewidth]{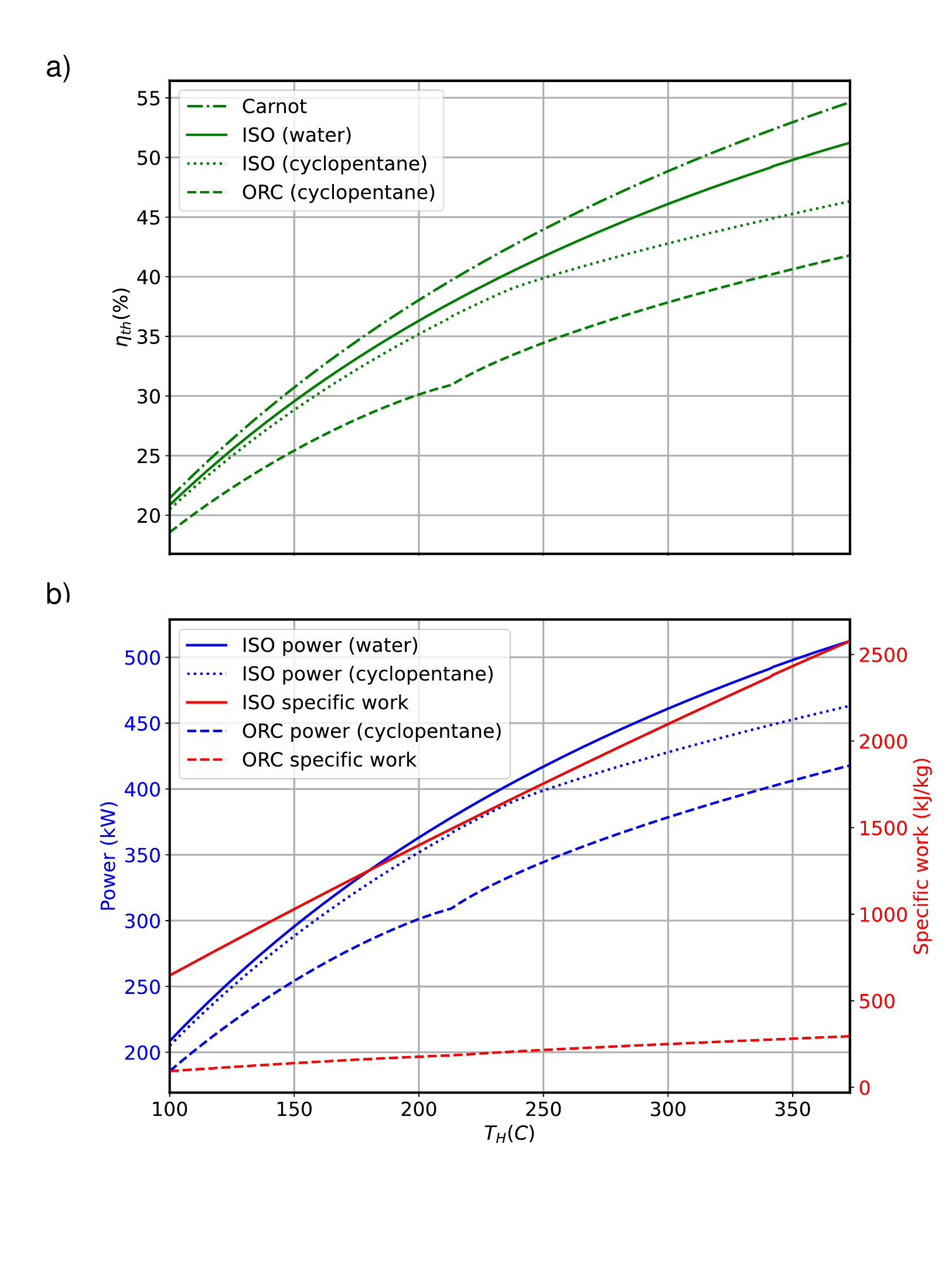}
\subsection*{Figure 10. Comparative analysis between the ISO cycle and ORC of efficiencies (a) and net output work (b).}
\subsection*{Table 1. ISO cycle (water) improvement}
\begin{tabular}{|c | c | c|} 
 \hline
 \textbf{$T_{H}$} & Power output improvement & Specific work improvement \\ [1ex] 
 \hline
 (${}^\circ$C) & (\%) & (\%)\\
 \hline
 100  & 12.3 & 591\\ [1ex]
 \hline
120  & 14.0 & 609\\ [1ex]
\hline
140  & 15.5 & 627\\ [1ex]
\hline
160  & 17.0 & 645\\ [1ex]
\hline
180  & 18.6 &  665\\ [1ex]
\hline
200  & 20.5 & 690\\ [1ex]
\hline
220  & 21.6 & 707 \\ [1ex]
\hline
240  & 21.0 & 708 \\ [1ex]
\hline
260  & 21.1 & 717\\ [1ex]
\hline
280  & 21.5 & 727\\ [1ex]
\hline
300  & 21.8 & 738\\ [1ex]
\hline
320  & 22.1 & 749\\ [1ex]
\hline
340  & 22.3 & 759\\ [1ex]
\hline
360  & 22.6 & 769\\ [1ex]
\hline
373  & 22.6 & 773\\ [1ex]

 \hline
\end{tabular}

\bigskip




Since the reaction turbine has yet to be built and experimented with, we consider turbine efficiencies in the range of 20-90\%, where the ISO turbine efficiency is defined by the ratio of the turbine output power to the WF expansion work. In this context, we assume the heat produced due to turbine irreversibilities is retained within the system. In contrast to the ORC turbine, where the generated heat cannot be utilized, the ISO cycle incorporates this thermal energy into the HTL stream, thereby augmenting the flow rate of the working fluid. This process mitigates the decline in work output due to turbine inefficiencies. Figure 11 illustrates the output power enhancement of the ISO cycle using water as the working fluid across the varying turbine efficiencies relative to the ORC. In this analysis, we consider a realistic ORC expander isentropic efficiency of 75\%, consistent with values used in past studies \cite{Bianchi2011}. The black curve in the figure denotes the ISO turbine efficiency for which the output power of the two cycles is equal. The figure indicates that at the considered ISO turbine efficiencies range, the ISO cycle potentially generates up to 41\% more power than the ORC-based heat engine and is superior to the ORC-based heat engines for ISO turbine efficiencies exceeding 65\% in the examined temperature range.

\noindent\includegraphics[width=0.85\linewidth]{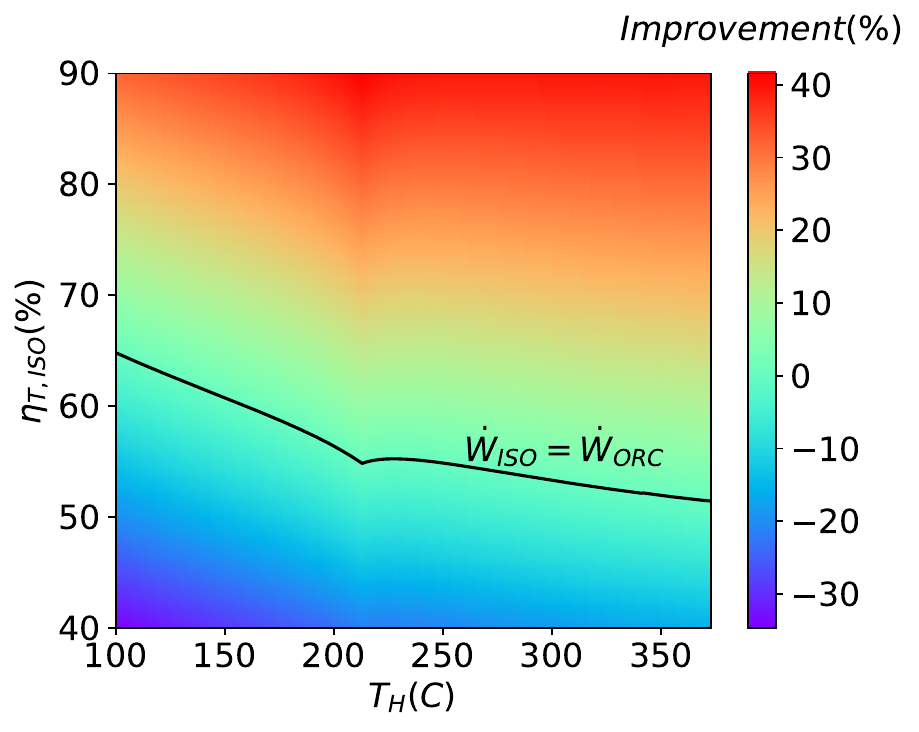}
\subsection*{Figure 11. ISO cycle power output improvement compared to the ORC as a function of the heat source temperature and the ISO turbine efficiency considering ORC turbine isentropic efficiency of 75\%. The black curve denotes the ISO turbine efficiency for which the ISO cycle and ORC produce equal power.}

When applied to ideal gas cycles, such as the Ericsson cycle, the proposed turbine may offer even higher-efficiency heat engines. There is no phase change in this cycle, and the gas undergoes isothermal compression, theoretically reaching Carnot's efficiency. Thus, the successful implementation of effective isothermal compression could pave the way for developing heat engines with efficiencies surpassing those achievable by the ISO cycle.

\subsection*{Conclusions}
This study introduces the concept of a supersonic two-phase nozzle as a key component of a novel thermodynamic cycle utilizing isothermal expansion. Experimental pressure measurements with air in water validated the supersonic conditions, while CFD allowed us to extract an upper bound for the polytropic index of air of $n < 1.052$. This index indicates near-isothermal expansion, enabling up to 71\% more work extraction than adiabatic expansion. Compared to other reported methods, such as flooded expansion, this method achieves a much higher liquid-to-working-fluid mass ratio (0.997 compared to 0.3). This significantly increases the heat transfer between the phases, enabling almost isothermal gas expansion. Furthermore, using the two-phase nozzle is an important step for increasing the pressure ratio, making it more suitable for various applications.

To leverage this nozzle for realizing a heat engine, we theoretically explored a reaction turbine, rotating due to thrust created by the nozzles. CFD analysis of the impeller shows hydraulic efficiency of  $\sim97\%$, supporting its efficiency potential compared to existing small-scale turbines.

The proposed heat engine, based on the novel cycle (termed ISO cycle), yielded superior thermodynamic performance compared to a heat engine based on a conventional ORC cycle and showed up to 22.6\% higher power output for heat source temperatures up to 373${}^\circ$C.

We broadened our simulation to encompass turbine efficiencies spanning from 40\% to 90\% and compared to ORC small-scale heat engine, which encompasses turbine isentropic efficiency of 75\%. This analysis revealed that the ISO cycle outperforms the ORC within the examined temperature range when turbine efficiency exceeds 65\%. \newline
Not only is it more efficient, but the ISO turbine also has the potential to be significantly smaller than its ORC counterpart. This is because the energy carrier in the ISO cycle is liquid, three orders of magnitude denser than gas, which is the typical energy carrier in ORC and other heat engines.


Future research should focus on building and experimentally validating the proposed cycle, optimizing its components, and exploring its potential applications in various energy systems.





\newpage


\section*{Experimental procedures}

\subsection*{Nozzle CFD analysis} \label{CFD nozzle analysis}
A 3D CFD simulation using the finite volume methods was performed on a nozzle using the commercial solver FLUENT 2024R1. The CFD model was validated by comparing the static pressure results with the experimental measurements along the nozzle. Due to the highly dispersed gas phase in the liquid flow and the high boundary area between phases, a 'Mixture' multiphase model was used. The 'Mixture' multiphase model is a simplified model that treats the mixture as a single continuum without sharp boundaries between the phases. The standard fluid dynamic governing equations are modified by introducing the void fraction parameter. No slip velocity between phases was assumed. 
A steady Reynolds Averaged Navier-Stokes (RANS) turbulence modeling approach was used to minimize simulation runtime. The RANS approach provides a solution for the time-averaged turbulent flow parameters. Although it eliminates the fluctuating terms, it is a powerful simplification that allows reasonably accurate results under reasonable computation costs. As a closure model, K-omega SST was compared with the K-epsilon models. Both models produced similar results.
The simulations were performed with several cell sizes to verify the independence of the results from cell size. Polyhidara cells were used. Prism cells were modeled for the wall treatment. Using the Effective Viscosity Ratio (EVR) parameter, a sufficient number of prism cells was assured, ensuring the actual boundary layer falls within the scope of the prism cells. 

\subsection*{Impeller CFD analysis} \label{CFD impeller analysis}
The impeller is designed and optimized using ANSYS 2024R1 software. For this purpose, a few ANSYS modules were used. We used Vista CPD to define the initial impeller geometry for specific operating conditions,  followed by BladeGen to create the final blade, shroud, and hub geometries. In the next step, the fast solver of VistaTF is applied to check operating conditions. Finally, Fluent's full CFD solution is done to evaluate impeller efficiency and $NPSH_r$. The fine mesh with sufficient boundary layer modeling was used, and three levels of mesh refinement validated the convergence. For all meshes, y+ was less than 1.

\subsection*{ISO cycle Simulation model}

In this subsection, we outline the thermodynamic framework applied to assess the ISO cycle.

For each temperature, we optimized the temperature $t_3$ for maximum output power and solved the following equations:
\begin{enumerate}
\item 
$\dot{w}_{exp} = t_7(s_7 - s_6) - (h_7-h_6)$
\item  
$\dot{q}_{in} = h_6-h_{11} + t_8(s_7 - s_6)$
\item 
$\dot{m}_{WF} = \frac{\dot{Q}_{in}}{\dot{q}_{in}} $
\end{enumerate}
where $s$, $h$, $t$, $\dot{q}$, $\dot{w}$, $\dot{V}$ and $\dot{m}$ are the specific entropy, enthalpy, temperature, heat, work, volumetric flow rate and mass flow rate, respectively, and the numbered subscripts denote the thermodynamic cycle points presented in Figure 8. $\dot{Q}_{in}$ is the rate heat transferred into the system.

For each heat source temperature, we optimize the nozzle's outlet temperature to achieve maximum net power output.

The thermal efficiency of the cycle is calculated as follows:
\begin{equation}
\eta_{th} = \frac{\dot{W}_{net}}{\dot{Q}_{in}},
\end{equation}

The net output work of the heat engine is calculated as:
\begin{equation}
\dot{W}_{net} = \dot{W}_{T} - \dot{W}_{p,WF} 
\end{equation}

where $\dot{W}_{T}$ and $\dot{W}_{p,WF}$ are the turbine's output work and the work consumed by the WF pump, respectively.

The turbine's output work is calculated as:
\begin{equation}  \label{turbine}
\dot{W}_{T} = \dot{W}_{exp} \times \eta_{T,ISO}
\end{equation}
where $\dot{W}_{exp}$ and $\eta_{T,ISO}$ are the expansion work and ISO turbine efficiency, respectively.

\section*{RESOURCE AVAILABILITY}


\subsection*{Lead contact}


Requests for further information and resources should be directed to and will be fulfilled by the lead contact, Dror Miron (drormiron@gmail.com).

\subsection*{Materials availability}


This study did not generate new materials.

\subsection*{Data and code availability}


\begin{itemize}
    \item Any information required to reanalyze the data reported in this paper is available from the lead contact upon request.    
\end{itemize}

\section*{ACKNOWLEDGMENTS}
The work of D. Miron was supported by the Israel Department of Energy (PhD fellowship) and by the Nancy and Stephen Grand Technion Energy Program (GTEP). The authors acknowledge the contribution of Matan Hameiry to this work.


\section*{AUTHOR CONTRIBUTIONS}


Conceptualization, D.M., Y.N, J.C., and C.R; methodology, D.M., Y.N, J.C., and C.R.; investigation, D.M., Y.N, N.F, A.S, and J.C.; visualization, D.M.; writing-–original draft, D.M., Y.N, J.C., and C.R.; writing-–review \& editing, D.M., Y.N, J.C., A.S, N.F, and C.R.; funding acquisition, C.R.; supervision, C.R.

\section*{DECLARATION OF INTERESTS}


All authors are employees and stakeholders of Lava Energy Ltd.
Carmel Rotschild is a founder of Lava Energy and a member of its scientific advisory board.
The authors have submitted the following relevant patents:
\begin{itemize}
\item	HEAT ENGINE WO2022/049573
\item	TWO PHASE HEAT ENGINE WO2023228173
\item	Heat engine using a liquid-vapor-phase-changing material  WO2024/028878
\end{itemize}

\bibliography{references}

\bigskip

\end{document}